\documentclass[preprint,showpacs,preprintnumbers,amsmath,amssymb]{revtex4}

\usepackage{graphicx}
\usepackage{dcolumn}
\usepackage{bm}

\renewcommand{\bar}[1]{\overline{#1}}

\begin{document}


\title{The Kaon Form Factor in the Light-Cone
Quark Model}

\author{Bo-Wen  Xiao}
\affiliation{Department of Physics, Peking University, Beijing
100871, China}
\author{Xin Qian}
\affiliation{Department of Physics, Peking University, Beijing
100871, China}
\author{Bo-Qiang Ma}%
\email{mabq@phy.pku.edu.cn}\altaffiliation{corresponding author.}
\affiliation{ CCAST (World Laboratory), P.O.~Box 8730, Beijing
100080, China\\
Department of Physics, Peking University, Beijing 100871, China}

\date{\today}

\begin{abstract}
   The electromagnetic form factor of the kaon meson is calculated in the
light-cone formalism of the relativistic constituent quark model.
The calculated $K^+$ form factor is consistent with almost all of
the available experimental data at low energy scale, while other
properties of kaon could also be interrelated in this
representation with reasonable parameters. Predictions of the form
factors for the charged and neutral kaons at higher energy scale
are also given, and we find non-zero $K^0$ form factor at $Q^2 \ne
0$ due to the mass difference between the strange and down quarks
inside $K^0$.
\end{abstract}

\pacs{14.40.AQ; 12.39.Ki; 13.40.Gp}


\vfill

{\centerline{Accepted for publication in Eur.Phys.J.A.}}


\maketitle


   The light-cone formalism \cite{Bro89, Lep80} provides a
convenient framework for the relativistic description of hadrons
in terms of quark and gluon degrees of freedom. The
electromagnetic form factor of pion has been studied and discussed
\cite{Ma93, Ma95, Cao97} in the light-cone formalism, which has
proved successful in explaining the experimental data. Similar to
the pion, the kaon is also composed by two quarks, but with
different quark masses. Therefore, it becomes a little more
complicated to obtain the light-cone wave function of the kaon and
to compute the kaon space-like form factor. However, unlike the
$\pi^0$, which has zero form factor due to its quark and antiquark
with opposite charges (i.e., a pair of quark and antiquark of the
same flavor), the $K^0$ form factor will be non-zero due to the
different contributions from the strange ($\bar{s}$) and down
($d$) quarks inside $K^0$. Thus measurements of the form factors
of the charged kaon ($K^{\pm}$) and neutral kaon ($K^0$ and
$\bar{K}^0$) will provide more information concerning the internal
structure of the mesons.

   In order to obtain the light-cone spin space wave function of the
kaon, we transform the ordinary instant-form SU(6) quark model
space wave function of the kaon into light-cone
dynamics \cite{Ma93, Ter76, Kar80, Chu88}. In the kaon rest frame
$(q_1+q_2=0)$, the instant-form spin space wave function of the
kaon is
\begin{equation}\label{instant-form}
   \chi_T=(\chi_1^{\uparrow}\chi_2^{\downarrow}-
   \chi_2^{\uparrow}\chi_1^{\downarrow})/\sqrt{2},
\end{equation}
in which $\chi_i^{\uparrow,\downarrow}$ is the two-component Pauli
spinor and the two quarks have 4-momentum
$q_1^{\mu}=(q_1^0,\mathbf{q})$ and
$q_2^{\mu}=(q_2^0,-\mathbf{q})$, with
$q_i^0=(m_i^2+{\bf q}^2)^{1/2}$, respectively. The instant-form spin
states $|J,s\rangle_T$ and the light-cone form spin states
$|J,\lambda\rangle_F$ are related by a Wigner rotation
$U^J$ \cite{Wig39}
\begin{equation}\label{Wigner}
  |J,\lambda\rangle_F=\sum_s U_{s\lambda}^J |J,s\rangle_T .
\end{equation}
   This rotation is called the Melosh rotation \cite{Mel74}
for spin-1/2 particles. Applying the transformation
Eq.~(\ref{Wigner}) on the both sides of Eq.~(\ref{instant-form}),
we can obtain the spin space wave function of the kaon in the
infinite-momentum frame. For the left side, i.e., the kaon, the
transformation is simple since the Wigner rotation is unity. For
the right side, i.e., two spin-1/2 partons, the instance-form and
light-front form spin states are related by the Melosh
transformation \cite{Wig39, Mel74, Ma91},
\begin{eqnarray}\label{Melosh}
    \chi_1^{\uparrow}(T)&=&\omega_1[(q_1^+
  +m_1)\chi_1^{\uparrow}(F)-q_1^R\chi_1^{\downarrow}(F)], \nonumber\\
    \chi_1^{\downarrow}(T)&=&\omega_1[(q_1^+
  +m_1)\chi_1^{\downarrow}(F)+q_1^L\chi_1^{\uparrow}(F)], \nonumber\\
    \chi_2^{\uparrow}(T)&=&\omega_2[(q_2^+
  +m_2)\chi_1^{\uparrow}(F)-q_2^R\chi_2^{\downarrow}(F)], \nonumber\\
    \chi_2^{\downarrow}(T)&=&\omega_2[(q_2^+
  +m_2)\chi_1^{\downarrow}(F)+q_2^L\chi_2^{\uparrow}(F)],
\end{eqnarray}
where $\omega_i=[2q_i^+(q_i^0 + m_i)]^{-1/2}$, $q_i^{R,L}=q_i^1\pm
q_i^2$, and $q_i^+=q_i^0+q_i^3$. Then we get the light-cone spin
wave function for the kaon,
\begin{equation}\label{spin}
\chi^K(x,{\bf k}_{\perp})=\sum_{\lambda_1,\lambda_2} C_0^F
(x,{\bf k}_{\perp},\lambda_1,\lambda_2)\chi_1^{\lambda_1}(F)\chi_2^{\lambda_2}(F),
\end{equation}
where the component coefficients $C_{J=0}^F
(x,{\bf k}_{\perp},\lambda_1,\lambda_2)$, when expressed in terms of the
instant-form momentum $q_{\mu}=(q^0,\mathbf{q})$, have the forms:
\begin{eqnarray}\label{coefficient}
C_0^F (x,{\bf k}_{\perp},\uparrow,\downarrow)&=&\omega_1
\omega_2[(q_1^+
+m_1)(q_2^+ +m_2)-{\bf q}_{\perp}^2]/\sqrt{2},\nonumber\\
 C_0^F (x,{\bf k}_{\perp},\downarrow,\uparrow)&=&-\omega_1 \omega_2[(q_1^+
+m_1)(q_2^+ +m_2)-{\bf q}_{\perp}^2]/\sqrt{2},\nonumber\\
C_0^F(x,{\bf k}_{\perp},\uparrow,\uparrow)&=&\omega_1
\omega_2[(q_1^+
+m_1)q_2^L-(q_2^+ +m_2)q_1^L]/\sqrt{2},\nonumber\\
C_0^F(x,{\bf k}_{\perp},\downarrow,\downarrow)&=&\omega_1
\omega_2[(q_1^+ +m_1)q_2^R-(q_2^+ +m_2)q_1^R]/\sqrt{2},
\end{eqnarray}
which satisfy the relation,
\begin{equation}\label{norm}
\sum_{\lambda_1,\lambda_2} C_0^F
(x,{\bf k}_{\perp},\lambda_1,\lambda_2)^\ast C_0^F
(x,{\bf k}_{\perp},\lambda_1,\lambda_2)=1.
\end{equation}
We can see that there are also two higher helicity
$(\lambda_1+\lambda_2 = \pm 1)$ components in the expression of
the light-cone spin wave function of the kaon besides the ordinary
helicity $(\lambda_1+\lambda_2 =0)$ components. Such higher helicity components \cite{Ma93, Ma95, Cao97}
come from the Melosh-rotation, and the same effect plays an important
role to understand the proton ``spin puzzle" in the nucleon case \cite{Ma91,Ma96}.

   Furthermore, we still have to know the space wave function.
Unfortunately, there is no exact solution of the Bethe-Salpeter
equation for the kaon at present. Approximately, we can adopt the
commonly used harmonic oscillator wave function instead,
\begin{equation}\label{harmonic}
  \varphi (q^2)=A \exp(-q^2/2{\beta}^2),
\end{equation}
which is a non-relativistic solution of the Bethe-Salpeter
equation in an instantaneous approximation in the rest frame for
meson \cite{EPT}. By assuming that the relation between the
instant-form momentum ${\bf q}=(q^3,{\bf q}_{\perp})$ and the light-cone
momentum $\underline{k}=(x,{\bf k}_{\perp})$ is by no means unique, and
according to the light-cone formalism, we construct models to
relate them. In this presentation, we adopt the
connection\cite{Ter76, Kar80, Chu88} in the light-front dynamics:
\begin{eqnarray}\label{light-cone}
          x_1M&=&q_1^0+q_1^3,\nonumber\\
          x_2M&=&q_2^0+q_2^3,\nonumber\\
          {\bf k}_{\perp}&=&{\bf q}_{\perp},
\end{eqnarray}
here $x_i$ $(i=1,2)$, with $x_1+x_2=1$, is the light-cone momentum
fraction of quark in the 2-particle Fock state. In the rest frame
$(q_1+q_2=0)$, from Eq.~(\ref{light-cone}) we can find that $M$
satisfies:
\begin{equation}\label{M-square1}
 M^2=\frac{m_1^2+{\bf k}^2_{\perp}} {x_1}+\frac{m_2^2+{\bf k}^2_{\perp}}{x_2} .
\end{equation}
If we let $x_1=x$, then we can get $x_2=1-x$. Then
Eq.~(\ref{M-square1}) can be written as follows:
\begin{equation}\label{M-square}
  M^2=\frac{m_1^2+{\bf k}^2_{\perp}} {x}+\frac{m_2^2+{\bf k}^2_{\perp}}{1-x}.
\end{equation}
From Eq.~(\ref{light-cone}) we can also obtain:
\begin{eqnarray}\label{Q1}
q_1^0&=&\frac{1}{2} M x+\frac{{\bf k}^2_{\perp}+m_1^2}{2xM},\nonumber\\
q_1^3&=&\frac{1}{2} M x-\frac{{\bf k}^2_{\perp}+m_1^2}{2xM},
\end{eqnarray}
\begin{eqnarray}\label{Q2}
q_2^0&=&\frac{1}{2} M (1-x)+\frac{{\bf k}^2_{\perp}+m_2^2}{2(1-x)M},\nonumber\\
q_2^3&=&\frac{1}{2} M (1-x)-\frac{{\bf
k}^2_{\perp}+m_2^2}{2(1-x)M}.
\end{eqnarray}
From Eq.~(\ref{Q1}) and Eq.~(\ref{Q2}) we can easily find that
$q_1^3=-q_2^3$. Thus we have:
\begin{eqnarray}\label{Q3}
q_1^+&=&xM,\nonumber\\
q_2^+&=&(1-x)M,
\end{eqnarray}
\begin{eqnarray}\label{Q4}
2q_1^+(q_1^0+m_1)&=&(xM+m_1)^2+{\bf k}^2_{\perp},\nonumber\\
2q_2^+(q_2^0+m_2)&=&[(1-x)M+m_2]^2+{\bf k}^2_{\perp}.
\end{eqnarray}
Then we can get:
\begin{equation}\label{Q-square}
  q^2=(q_1)^2=(q_2)^2=\frac{1}{4}
  M^2+\frac{(m_1^2-m_2^2)^2}{4M^2}-\frac{1}{2}(m_1^2+m_2^2),
\end{equation}
where
$$M^2=\frac{m_1^2+{\bf k}^2_{\perp}}
{x}+\frac{m_2^2+{\bf k}^2_{\perp}}{1-x}. $$
There is still another way to obtain Eq.~(\ref{Q-square}).
Brodsky-Huang-Lepage suggested a connection between the equal-time
wave function in the rest frame and the light-cone wave function
by equating the off-shell propagator
$\varepsilon=M^2-{({\sum}_{i=1}^nk_i)}^2$ in the two
frames \cite{BHL81}:
\begin{displaymath}
\varepsilon=\left\{\begin{array}{ll}
M^2-(\sum_{i=1}^n q_i^0)^2,
&\sum_{i=1}^nq_i=0, ~{\mathrm [C.M]}\\
M^2-\sum_{i=1}^n \frac{{\bf k}_{\perp i}^2+m_i^2}{x_i},
&\sum_{i=1}^n {\bf k}_{\perp i}=0,~ \sum_{i=1}^n x_i=1.~
{\mathrm [L.C]}
\end{array}\right.
\end{displaymath}
From the equation above, for two-particle systems one can get:
\begin{equation}\label{Q-square1}
  q^2=\frac{1}{4}(\frac{{\bf k}_{\perp}^2+m_1^2}{x}+\frac{{\bf k}_{\perp}^2+m_2^2}{1-x})
  +\frac{(m_1^2-m_2^2)^2}{4(\frac{{\bf k}_{\perp}^2+m_1^2}{x}+\frac{{\bf k}_{\perp}^2+m_2^2}{1-x})}
  -\frac{1}{2}(m_1^2+m_2^2).
\end{equation}

   Obviously, Eq.~(\ref{Q-square}) and Eq.~(\ref{Q-square1}) are
the same, which is to say that although having employed different
assumptions at the beginning, we obtain the same result of $q^2$
at last. This may indicate that the model that we have established
is self-explained.

   By adopting the Brodsky-Huang-Lepage (BHL)
prescription \cite{Lep80}, we can obtain:
\begin{equation}\label{BHL}
  \varphi_{\mathrm{BHL}}=A_0 \exp[-\frac{\frac{{\bf k}_{\perp}^2+m_1^2}{x}+\frac{{\bf k}_{\perp}^2+m_2^2}{1-x}}{8{\beta}^2}-
  \frac{(m_1^2-m_2^2)^2}{8{\beta}^2(\frac{{\bf k}_{\perp}^2+m_1^2}{x}+\frac{{\bf k}_{\perp}^2+m_2^2}{1-x})}],
\end{equation}
in which we let $A_0=A \exp(\frac{m_1^2+m_2^2}{4{\beta}^2})$. The
contributions from non-zero transversal momentum $\left|{\bf
k}_{\perp}\right|$ in the end-point $x \to 0$ and $x \to 1$
regions are highly suppressed by the exponential fall-off, so this
wave function provides an automatic cut-off on $\left|{\bf
k}_{\perp}\right|$. This feature is introduced via the BHL
prescription which relies on the free light-cone hamiltonian.

   Therefore, the light-cone wave function for the kaon can be
written as following:
\begin{equation}\label{wave}
 \psi=\varphi_{\mathrm{BHL}} \chi^K(x,{\bf k}_{\perp}),
\end{equation}
in which the parameters are the normalization constant $A_0$, the
harmonic scale $\beta$ and the quark masses $m_1$ and $m_2$. Thus
we employ the following four constraints to adjust those above
four parameters:

   1. The normalization condition:
\begin{equation}\label{normalization}
\int \frac{{\mathrm{d}} ^2 {\bf k}_{\perp}{\mathrm{d}} x}{16{\pi}^3} {\psi}^{\ast}\psi=\int
\frac{{\mathrm{d}} ^2{\bf k}_{\perp}{\mathrm{d}} x}{16{\pi}^3} {\varphi}_{\mathrm{BHL}}^{\ast}
{\varphi}_{\mathrm{BHL}}=1,
\end{equation}
which is essentially a valence quark dominance
assumption \cite{Ma93}.

   2. The weak decay constant $f_K=113.4$ {MeV} is defined \cite{decay, Buc95}
from $K \to \mu \, \nu $ decay, thus one obtains:
\begin{equation}\label{decay}
  \int_{0}^{1} {\mathrm{d}} x \int \frac{{\mathrm{d}} ^2{\bf k}_{\perp}}{16{\pi}^3}
  \frac{(k_1^++m_1)(k_2^++m_2)-{{\bf k}_{\perp}}^2}{{[(k_1^++m_1)^2+{{\bf k}_{\perp}}^2]}^{1/2}{[(k_2^++m_2)^2+{{\bf k}_{\perp}}^2]}^{1/2}}
  {\varphi}_{\mathrm{BHL}}=\frac {f_K}{2\sqrt{3}}.
\end{equation}

   3. The charged mean square radius of $K^+$ is defined as:
\begin{equation}\label{K+}
\langle r_{K^+}^2 \rangle =-6\frac{\partial F_{K^+}(Q^2)}{\partial
Q^2}|_{Q^2=0}.
\end{equation}
We can find the experimental value of $\langle r_{K^\pm}^2
\rangle=0.34\pm 0.05$ {fm}$^2 $ \cite{data1}, and $\langle
r_{K^-}^2 \rangle=0.28\pm 0.05$ {fm}$^2 $ \cite{data2}.

   4. The charged mean square radius of $K^0$ is defined as:
\begin{equation}\label{K0}
\langle r_{K^0}^2 \rangle =-6\frac{\partial F_{K^0}(Q^2)}{\partial
Q^2}|_{Q^2=0}.
\end{equation}
We can find the experimental value of $\langle r_{K^0}^2
\rangle=-0.054\pm 0.026$ {fm}$^2$ \cite{data3}.

   Therefore, we can obtain $m_1=500$ {MeV} (e.g., the strange quark),
$m_2=250$ {MeV} (e.g., the up quark or the down quark, assuming
$m_u=m_d$ ), $\beta =393$ {MeV} and $A_0=0.0742$. It is
interesting to notice that the masses of the strange quarks and
the light-flavor quarks from the above constraints are just in the
correct range of the constituent quark masses from more general
considerations.

   Reversely, we can compute the value of $f_K$, $\langle r_{K^+}^2 \rangle$, and $\langle
r_{K^0}^2 \rangle$ by using the four parameters above:
\begin{eqnarray}\label{result}
f_K=113.3~ \mathrm{MeV},\nonumber\\
\langle r_{K^+}^2 \rangle=0.30~ {\mathrm{fm}}^2,\nonumber\\
\langle r_{K^0}^2 \rangle=-0.055~ {\mathrm{fm}}^2.
\end{eqnarray}
   The results fit the experimental values well. Naturally, the form
factor results emerging from this assumption are in quite good
agreement with the experimental data. Moreover, the values of the
parameters($m_1$, $m_2$, $\beta$) are compatible with other quark
models \cite{Choi98, Buc95, Guo91, Simula}.

   Since the Wigner rotation relating spin state in different
frames is unity under kinetic Lorentz transformation in the
light-cone formalism, the spin structures of hadrons are the same
in different frames related by Lorentz transformation. Therefore,
we can calculate the electromagnetic form factor from the
Drell-Yan-West formula \cite{DYW} by using the light-cone
formalism,
\begin{equation}\label{Drell}
  F(Q^2)=\sum_{n,
  {\lambda}_i}\sum_{j}e_j\int[{\mathrm{d}} x][{\mathrm{d}}^2
  {\bf k}_{\perp}]{\psi}_n^{\ast}(x_i,{\bf k}_{\perp i},{\lambda}_i)
  {\psi}_n(x_i,{\bf k}_{\perp i}^{\prime},{\lambda}_i),
\end{equation}
where ${\bf k}_{\perp i}^{\prime}={\bf k}_{\perp i}-x_i {\bf q}_{\perp}+{\bf q}_{\perp}$
for the struck quark, ${\bf k}_{\perp i}^{\prime}={\bf k}_{\perp i}
-x_i{\bf q}_{\perp}$ for the spectator quarks,
$[{\mathrm{d}}^2{\bf k}_{\perp}]={{\mathrm{d}}^2{\bf k}_{\perp}}/{16{\pi}^3}$, $e_j$ is the
electric charge of the struck quark, and the virtual photon
momentum $q_{\mu}$ is specified with $q^+=0$ to eliminate the
Z-graph contributions \cite{Bro89, Lep80, Ma89}. Other choice of
$q_{\mu}$ will cause contributions from Z-graphs, and it should
give the same result as that in the case of $q^+=0$ if all the
graphs are taken into account \cite{Saw92}. In the light-cone
formalism, there is a relation between $Q^2$ and
${\bf q}_{\perp}^2$:
\begin{equation}\label{q=Q}
 -Q^2=q^2=q^+q^--{\bf q}_{\perp}^2.
\end{equation}
Since $q^+=0$, then according to Eq.~(\ref{q=Q}), one can easily
get ${\bf q}_{\perp}^2=Q^2$.

   Because $K^+=u\bar{s}$ and $K^-=s\bar{u}$, one can find that
$F_{K^+}(Q^2)=-F_{K^-}(Q^2)$ according to Eq.~(\ref{Drell}).
Thereby, we just need to calculate the $K^+$ form factor
\begin{eqnarray}\label{Kaon+}
F_{K^+}(Q^2)=\frac{1}{3}e\int {\mathrm{d}} x\frac{{\mathrm{d}}^2{\bf k}_{\perp}}{16{\pi}^3}
{\cal{M}}_1 {\varphi}^{\ast}(x,{\bf k}_{\perp},m_1,m_2)
\varphi(x,{\bf k}_{\perp}^{\prime},m_1,m_2) \nonumber\\
+\frac{2}{3}e\int {\mathrm{d}} x\frac{{\mathrm{d}}^2{\bf k}_{\perp}}{16{\pi}^3}{\cal{M}}_2
{\varphi}^{\ast}(x,{\bf k}_{\perp},m_2,m_1)
\varphi(x,{\bf k}_{\perp}^{\prime},m_2,m_1),
\end{eqnarray}
where ${\bf k}_{\perp}^{\prime}={\bf k}_{\perp}+(1-x){\bf q}_{\perp}$ is the
internal quark transverse momentum of the struck kaon in the
center of mass frame, and
\begin{eqnarray}\label{M1M2}
{\cal{M}}_1&=&\frac{(a_1a_2-{\bf
k}^2_{\perp})(a_1^{\prime}a_2^{\prime}-{{\bf
k}_{\perp}^{\prime}}^2)+ (a_1+a_2)(a_1^{\prime}+a_2^{\prime}){\bf
k}_{\perp} \cdot {\bf k}_{\perp}^{\prime}}{{[(a_1^2+{\bf
k}^2_{\perp})(a_2^2+{\bf k}^2_{\perp})
({a_1^{\prime}}^2+{{\bf k}_{\perp}^{\prime}}^2)({a_2^{\prime}}^2+{{\bf k}_{\perp}^{\prime}}^2)]}^{1/2}},\\
{\cal{M}}_2&=& \frac{(b_1b_2-{\bf
k}^2_{\perp})(b_1^{\prime}b_2^{\prime}-{{\bf
k}_{\perp}^{\prime}}^2)+ (b_1+b_2)(b_1^{\prime}+b_2^{\prime}){\bf
k}_{\perp} \cdot {\bf k}_{\perp}^{\prime}}{{[(b_1^2+{\bf
k}^2_{\perp})(b_2^2+{\bf k}^2_{\perp}) ({b_1^{\prime}}^2+{{\bf
k}_{\perp}^{\prime}}^2)({b_2^{\prime}}^2+{{\bf
k}_{\perp}^{\prime}}^2)]}^{1/2}},
\end{eqnarray}
in which
\begin{eqnarray}
\label{explain}
a_1&=&xM_a+m_1,  ~~  m_1=m_{\bar{s}}=500~{\mathrm{MeV}}\nonumber\\
a_2&=&(1-x)M_a+m_2,  ~~  m_2=m_u=250~{\mathrm{MeV}}\nonumber\\
a_1^{\prime}&=&xM_a^{\prime}+m_1,\nonumber\\
a_2^{\prime}&=&(1-x)M_a^{\prime}+m_2,\nonumber\\
b_1&=&xM_b+m_2,\nonumber\\
b_2&=&(1-x)M_b+m_1,\nonumber\\
b_1^{\prime}&=&xM_b^{\prime}+m_2,\nonumber\\
b_2^{\prime}&=&(1-x)M_b^{\prime}+m_1,\nonumber\\
\end{eqnarray}
and in which
\begin{eqnarray}\label{explain1}
 M^2_a=\frac{m_1^2+{\bf k}^2_{\perp}} {x}+\frac{m_2^2+{\bf k}^2_{\perp}}{1-x},\nonumber\\
 M^{\prime 2}_a=\frac{m_1^2+{\bf k}^{\prime 2}_{\perp}} {x}+\frac{m_2^2+{\bf k}^{\prime 2}_{\perp}}{1-x},\nonumber\\
 M^2_b=\frac{m_2^2+{\bf k}^2_{\perp}} {x}+\frac{m_1^2+{\bf k}^2_{\perp}}{1-x},\nonumber\\
 M^{\prime 2}_b=\frac{m_2^2+{\bf k}^{\prime 2}_{\perp}} {x}+\frac{m_1^2+{\bf k}^{\prime 2}_{\perp}}{1-x}.\nonumber\\
\end{eqnarray}

\begin{figure}[htb]
\includegraphics{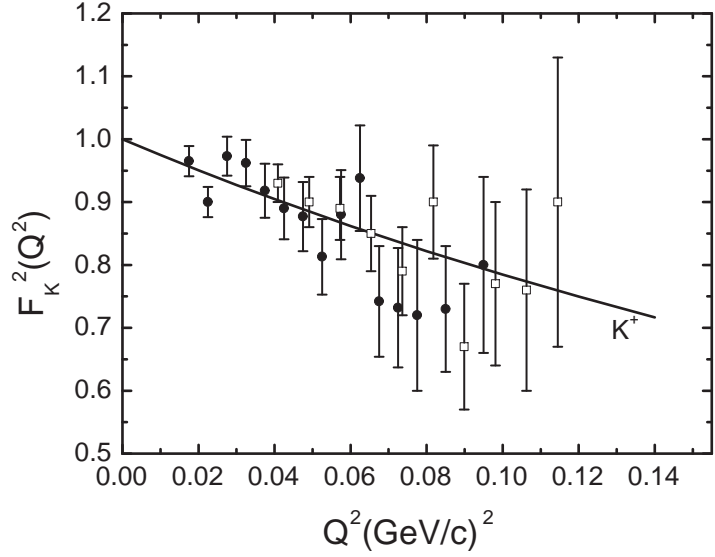}
\caption[*]{
\label{low1} The $K^+$ form factor calculated with the wave
function in the BHL prescription at low $Q^2$ compared with the
experimental data. The data are taken from Refs.~\cite{data1} and
\cite{data2}.}
\end{figure}

   For the same reason, since $K^0=d\bar{s}$ and
$\bar{K}^0=s\bar{d}$, one can also find that
$F_{K^0}(Q^2)=-F_{\bar{K}^0}(Q^2)$ according to Eq.~(\ref{Drell}).
Thereby, we just need to calculate the $K^0$ form factor,
\begin{eqnarray}\label{Kaon0}
F_{K^0}(Q^2)=\frac{1}{3}e\int {\mathrm{d}} x\frac{{\mathrm{d}}^2{\bf k}_{\perp}}{16{\pi}^3}
{\cal{M}}_1{\varphi}^{\ast}(x,{\bf k}_{\perp},m_1,m_2)
\varphi(x,{\bf k}_{\perp}^{\prime},m_1,m_2) \nonumber\\
-\frac{1}{3}e\int {\mathrm{d}}x\frac{{\mathrm{d}}^2{\bf k}_{\perp}}{16{\pi}^3}
{\cal{M}}_2{\varphi}^{\ast}(x,{\bf k}_{\perp},m_2,m_1)
\varphi(x,{\bf k}_{\perp}^{\prime},m_2,m_1).
\end{eqnarray}
The definitions of the ${\cal{M}}_1$ and ${\cal{M}}_2$ are the same
with the definitions in the calculation of $K^+$ form factor which
we have given above.

   Because of the existence of the mass difference between the down
quark and the strange quark, we find that $F_{K^0}(Q^2)\ne 0$ at
$Q^2 \ne 0$. In comparison, we may note that
$F_{{\pi}^0}(Q^2)\equiv 0$ because the ${\pi^0}$ has zero charge
and equal quark masses. Naturally, we can figure out that the
non-zero $K^0$ form factor is strongly dependent on the value of
the mass difference between the strange quark and the down quark.
This aspect is useful to reveal the different contributions from
the strange and down quarks inside $K^0$.

\begin{figure}[htb]
\includegraphics{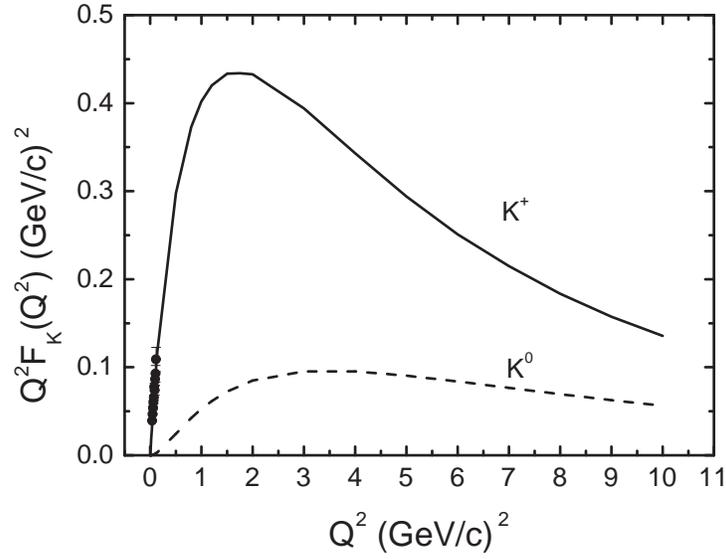}
\caption[*]{
\label{high} Theoretical electromagnetic form factor of the $K^+$
and $K^0$, represented by the solid and dashed lines respectively.
The data for the $K^+$ are taken from Ref.~\cite{data2}.}
\end{figure}

   Fig.~1 indicates that in the case of low $Q^2$, the theoretical
values of the $K^+$ form factor fit the experimental data very
well. The same model can also provide very good description of the
charged pion form factor for $Q^2 \leq 2$ (GeV/c)$^2$ \cite{Ma93}.
Because we are in lack of the experimental data for the $K^+$ in
the higher energy scale, we give the predictions of the
$Q^2F_{K}(Q^2)$ values for the $K^+$ and $K^0$ in Fig.~2. Since
the $K^0$ has zero charge, we can see that its electromagnetic
form factor is much less than the form factor of $K^+$. Similar
prediction has been also given in \cite{Choi98,Simula}. Thus it
needs high precision to measure the $K^0$ form factor
experimentally since the electro-production cross section is
small.

   It is necessary to point out that this work should be considered
as a light-cone version of the relativistic constituent quark
model \cite{Ma93,LCQM}, and it should be only valid in the low
energy scale of about $Q^2\leq 2$ (GeV/c)$^2$. Similar works have
been also developed in the systematic studies of mesons and
baryons \cite{Simula}.  It is different from the light-cone
perturbative QCD approach \cite{Lep80}, which is applicable at the
high energy scale of $Q^2
> 2$ (GeV/c)$^2$. The reason is that the hard gluon exchanges
between the quark-antiquark of the meson should be considered at
high $Q^2$, and this feature is incorporated in the light-cone
perturbative QCD approach. An ordinary input wave function may
contain uncertainties which invalid the prediction at high $Q^2$
in the constituent quark model framework. If the constituent quark
model prediction is happened to work at a higher $Q^2$, it might
be by chance or may imply a reasonable input wave function that
contains some features simulating the hard gluon exchanges. So the
agreement of a constituent quark model prediction with the
experiments might be served as a support of the applicability of
the input wave function from low energy scale to a somewhat higher
scale.

   In summary, we calculated the electromagnetic form
factor of the kaon by adopting the light-cone formalism of the
relativistic constituent quark model. By adjusting the parameters
through the experimental values of weak decay constant and charged
mean square radius, the model can give a good fit to the available
experimental values of kaon form factors. We also predicted the
form factors for both charged and neutral kaons, $K^{\pm}$ and
$K^0$. We expect the predictions to be valid when $Q^2 \leq 2$
(GeV/c)$^2$ at a low energy scale. A non-zero form factor of $K^0$
is predicted at $Q^2 \ne 0$, and it will be useful to reveal the
different contributions from strange and down quarks inside $K^0$.

{\bf Acknowledgments:} This work is partially supported by
National Natural Science Foundation of China under Grant Numbers 19975052, 10025523,
and 90103007.
It is also supported by
Hui-Chun Chin and Tsung-Dao Lee Chinese Undergraduate Research Endowment (CURE)
at Peking University.


\end{document}